\documentclass[preprint]{ptptex}

\title{
Modified Coulomb potential with virtual photons following a canonical distribution
}

\author{
Kohzo \textsc{Nishida}%
\footnote{E-mail: EZF01671@nifty.com} 
}

\inst{
Department of Physics, Kyoto Sangyo University, Kyoto 603-8555, Japan 
}

\abst{
The need for a cutoff in the Lamb shift calculation suggests that high-energy virtual photons do not interact with real particles.
In this paper, we assume that the creation of virtual photons follows a canonical distribution.
As a result, the Coulomb potential is modified to $V(r)=- ze^2\arctan (m_c r)/(2\pi^2 r)$, and 
the zero-point energy density of the electromagnetic field becomes finite.
}

\usepackage[dvipdfmx]{graphicx}
\begin{document}

\maketitle

\section{Introduction}
In quantum field theory, cutoffs are often introduced when calculating physical quantities.
Let us show this in the Lamb shift\cite{1,2,3,4,5} calculation.
The Lamb shift $\Delta E$ between the $2s_{1/2}$ and $2p_{1/2}$ levels is known to be\cite{6}
\begin{equation}
\label{eq:1}
\Delta E = \frac{\alpha^5 m_e}{6\pi} \int^{m_e}_{1/a_0} \frac{d p}{p}
= \frac{\alpha^5 m_e}{6\pi}  \ln \frac{1}{\alpha},
\end{equation}
where $m_e$ is the mass of the electron, $\alpha$ is the fine structure constant, and $a_0=1/(m_e\alpha)$ is the Bohr radius.
The energy integral of the virtual photon, $\int d p/p$, is stops counting the photons when their wavelength gets bigger than the size of the atom, $a_0$.  
On the short wavelength side, this integral stop counting the photons when their wavelength gets shorter than the Compton wavelength. 
That is, the cutoff $m_e$ is introduced in the Lamb shift calculation.
Renormalization by cutoff is also an operation that does not count virtual particles with energy larger than the cutoff.
Thus, the agreement between the experimental value and the theoretical value introducing the cutoff suggests that 
high-energy virtual photons do not virtually affect real particles.

Now, we propose a new approximate calculation method for integrals with cutoff.
(\ref{eq:1}) can be approximated by a suppression factor $e^{-p/m_e}$ as 
\begin{equation}
\label{eq:2}
\Delta E \simeq \frac{\alpha^5 m_e}{6\pi} \int^{\infty}_{1/a_0} \frac{d p}{p} e^{-p/m_e}
\simeq  \frac{\alpha^5 m_e}{6\pi}  \left ( \ln \frac{1}{\alpha} -\gamma + \pi \alpha \right ),
\end{equation}
where we used the integral formula
\begin{equation}
\label{eq:3}
\int^{\infty}_{x} \frac{e^{-p} }{p}d p = \Gamma(0, x) = (-\log(x) -\gamma) + x-\frac{x^2}{4} + \cdots,
\end{equation}
where $\Gamma(s, x)$ is the upper incomplete gamma function, and 
$\gamma=0.57721\cdots$ is  Euler's constant.
In addition to $\int d p/p$, we can approximate integrals
\begin{equation}
\label{eq:4}
I = \int^{\Lambda}_0 p^n dk = \frac{1}{n+1}\Lambda^{n+1}
\,\,\,\,\,\,
(n=0,1,2\cdots)
\end{equation}
that appears in the loop integral as follows when  $n$ is small:
\begin{equation}
\label{eq:5}
I \simeq \int^{\infty}_0 p^n e^{-p/\Lambda} d p = n! \Lambda^{n+1}.
\end{equation}
Thus, we can approximate an integral with a cutoff $\Lambda$ as an integral with an infinite integral range multiplied by the suppression factor $e^{-p/\Lambda}$.

The suppression factor $e^{-p/m_c}$ has the same form as a Boltzmann factor $e^{-\beta E}$.
If the suppressor is a Boltzmann factor, it means that the creation of virtual photons follows a canonical distribution\cite{7}.
In this paper, we investigate a model in which the creation of virtual photons 
follows a relativistic canonical distribution.
As a result, we demonstrate that the Coulomb potential is modified,
and the zero-point energy density of the electromagnetic field becomes finite.
The modified Coulomb potential has interesting properties that it becomes the ordinary Coulomb potential at a long distance and becomes finite at $r = 0$.

\section{Modified Coulomb potential}
We propose a new electromagnetic field operator with a probability density $\rho(\boldsymbol{p}, \bar{u})$, which is the c-number,
\begin{equation}
\label{eq:6}
A_\mu(x) 
=
\int \frac{d^3p} { \sqrt{(2\pi)^3 2\omega(\boldsymbol{p})}} 
\{
 a_\mu(\boldsymbol{p})e^{-ipx} 
+
a_\mu^{\dagger}(\boldsymbol{p}) e^{ipx}
\}
 \rho(\boldsymbol{p}, \bar{u})^{1/2},
\end{equation}
where we work in the Feynman gauge.
The coefficients of the expansion $a(\boldsymbol{p})$ and $a^\dagger(\boldsymbol{p})$ satisfy the canonical commutation relations
\begin{equation}
\label{eq:7}
[a_\mu(\boldsymbol{p}), a_\nu^\dagger (\boldsymbol{q})] = g_{\mu\nu}\delta^{3}(\boldsymbol{p}-\boldsymbol{q}), \,\,\,\,
[a_\mu(\boldsymbol{p}), a_\nu(\boldsymbol{q})] = 
[a_\mu^\dagger(\boldsymbol{p}), a_\nu^\dagger (\boldsymbol{q})] = 0,
\end{equation}
\begin{equation}
\label{eq:8}
a_\mu(\boldsymbol{p}) = 
\sum_{\lambda = 0}^{3} \varepsilon_\mu^\lambda(\boldsymbol{p}) a_{\boldsymbol{p}}^\lambda,
 \,\,\,\,\,\,
[a^\lambda_{\boldsymbol{p}}, a^{\lambda'\dagger}_{\boldsymbol{q}}] = g^{\lambda\lambda'}\delta^{3}(\boldsymbol{p}-\boldsymbol{q}).
\end{equation}
We assume that the probability density $\rho(\boldsymbol{p}, \bar{u})$ is a relativistic canonical distribution
\begin{equation}
\label{eq:9}
\rho(\boldsymbol{p}, \bar{u}) 
\equiv
\frac{e^{-|\bar{u}_\mu p^\mu(\boldsymbol{p})|/m_c }  } {Z}, 
\end{equation}
with
\begin{eqnarray}
\label{eq:10}
Z 
\equiv
 \iint d^3p d^3x e^{-|u_\mu p^\mu(\boldsymbol{p})|/m_c }
,
 \,\,\,\,\,\,
p^\mu(\boldsymbol{p}) 
\equiv
 \left( \omega(\boldsymbol{p}),\boldsymbol{p} \right),
\end{eqnarray}
where $m_c$ is a cutoff.
$Z$ is the state sum,
which is Lorentz invariant because $d^3p d^3x$ is Lorentz invariant.
$\bar{u}_\mu$ is an average of four-velocities of photons created by the potential $A_\mu(x) $.
The spatial distribution of matter in the universe is homogeneous and isotropic, 
which means that the average four-velocities of the photons, $\bar{u}^\mu$ can be written in any coordinate system as 
\begin{equation}
\label{eq:11}
\bar{u}_\mu = (1,0,0,0),
\end{equation}
because the spatial component is canceled with plus and minus appearing equally.
Thus, in any coordinate system, we can always rewrite the relativistic canonical distribution to
\begin{equation}
\label{eq:12}
\frac{e^{-|u_\mu p^\mu(\boldsymbol{p})|/m_c } }{Z}
=
\frac{ e^{-\omega(\boldsymbol{p})/m_c } }{Z}.
\end{equation}

Using (\ref{eq:6}), we obtain the photon propagator,  
\begin{eqnarray}
\label{eq:13}
\lefteqn{ D_{\mu\nu}(x,y,\bar{u}) } \nonumber \\
&=&
 i<0|T(A_\mu(x,u)A_\nu(y,u))|0> \nonumber \\
&=& 
i\int \frac{d^3 p}{\sqrt{(2\pi)^3 2\omega(\boldsymbol{p})}} \int \frac{d^3 q}{\sqrt{(2\pi)^3 2\omega(\mbox{\boldmath $q$})}} 
e^{-|\bar{u}_\mu p^\mu(\boldsymbol{p})|/(2m_c) }
e^{-|\bar{u}_\mu q^\mu(\boldsymbol{q})|/(2m_c) }
 \nonumber \\
&& 
\times[ <0'|\theta(x^0-y^0) a_\mu(\boldsymbol{p}) e^{-ipx} a_\nu^{\dagger}(\mbox{\boldmath $q$}) e^{iqy}
+\theta(y^0-x^0) a_\nu(\mbox{\boldmath $q$}) e^{-iqy} a_\mu^{\dagger}(\boldsymbol{p}) e^{ipx} |0'>] \nonumber \\
&=&
 i\int \frac{d^3 p}{(2\pi)^3 2\omega(\boldsymbol{p})} 
e^{-|\bar{u}_\mu p^\mu(\boldsymbol{p})|/m_c }
\nonumber \\
&&
\times g_{\mu\nu}\{ \theta(x^0-y^0) e^{-ip(x-y)} + \theta(y^0-x^0) e^{ip(x-y)} \} |_{p^{0} = \omega(\boldsymbol{p}) }
\end{eqnarray}
where we redefine the vacuum as
\begin{equation}
\label{eq:14}
|0'>\equiv \frac{1}{ \sqrt{Z}}|0>, \,\,\,\,\,\, <0'|0'>= 1.
\end{equation}
(\ref{eq:13}) can be rewritten as
\begin{eqnarray}
\label{eq:15}
D_{\mu\nu}(x,y,\bar{u})
&=&
 \int \frac{d^4 p}{(2\pi)^4} e^{-ip(x-y)} 
\frac{-g_{\mu\nu} }{p^2 - i\epsilon} 
e^{-|\bar{u}_\mu p^\mu(\boldsymbol{p})|/m_c },
\end{eqnarray}
where we use $\omega(-\boldsymbol{p}) = \omega(\boldsymbol{p})$. 
We obtain the photon propagator in momentum-space:
\begin{eqnarray}
\label{eq:16}
iD_{\mu\nu}(p, \bar{u}) 
&\equiv&
\int d^4 x e^{ipx} iD_{\mu\nu}(x,0,\bar{u})
\nonumber \\
&=&
\frac{ -i g_{\mu\nu}}{ p^2 -i\epsilon}
e^{-|\bar{u}_\mu q^\mu(\boldsymbol{p})|/m_c }.
\end{eqnarray}

Equation (\ref{eq:16}) shows that the Coulomb potential takes the following form\cite{8}
\begin{eqnarray}
\label{eq:17}
V(r)
&=&
 -\frac{ze^2}{(2\pi)^3} \int d^3 p e^{-i \boldsymbol{p}\cdot\boldsymbol{r} }  \frac{1}{|\boldsymbol{p}|^2}
e^{-|\bar{u}_\mu q^\mu(\boldsymbol{p})|/m_c }
\nonumber \\
&=&
-\frac{ze^2}{(2\pi)^3}
 \int d^3 p
 e^{-i \boldsymbol{p}\cdot\boldsymbol{r} }  
\frac{1}{|\boldsymbol{p}|^2}
e^{-|\boldsymbol{p}|/m_c },
\end{eqnarray}
where $z$ is the atomic number,
and we used (\ref{eq:12}).
This polar coordinate expression is 
\begin{eqnarray}
\label{eq:18}
V(r)
&=&
-\frac{ze^2}{(2\pi)^3}
\int_0^\infty dp \int_0^\pi d \theta \int_0^{2\pi} d\phi p^2 \sin \theta e^{-ipr \cos \theta} 
\frac{1}{p^2}
e^{-p/m_c }
\nonumber \\
&=&
-\frac{ze^2}{(2\pi)^3}
\frac{4\pi}{r}
\int_0^\infty dp \frac{\sin pr}{p} 
e^{-p/m_c },
\end{eqnarray}
where $p=|\boldsymbol{p}|$.
Using the integral formula
\begin{equation}
\label{eq:19}
\int_0^\infty  dx \frac{\sin rx}{x} 
e^{-x/m }
=
\arctan (mr),
\end{equation}
we finally obtain the modified Coulomb potential:
\begin{equation}
\label{eq:20}
V(r)
=
-\frac{ze^2}{2\pi^2 r} \arctan (m_c r).
\end{equation}
\begin{figure}[b]
\begin{center}
\includegraphics[width=90mm]{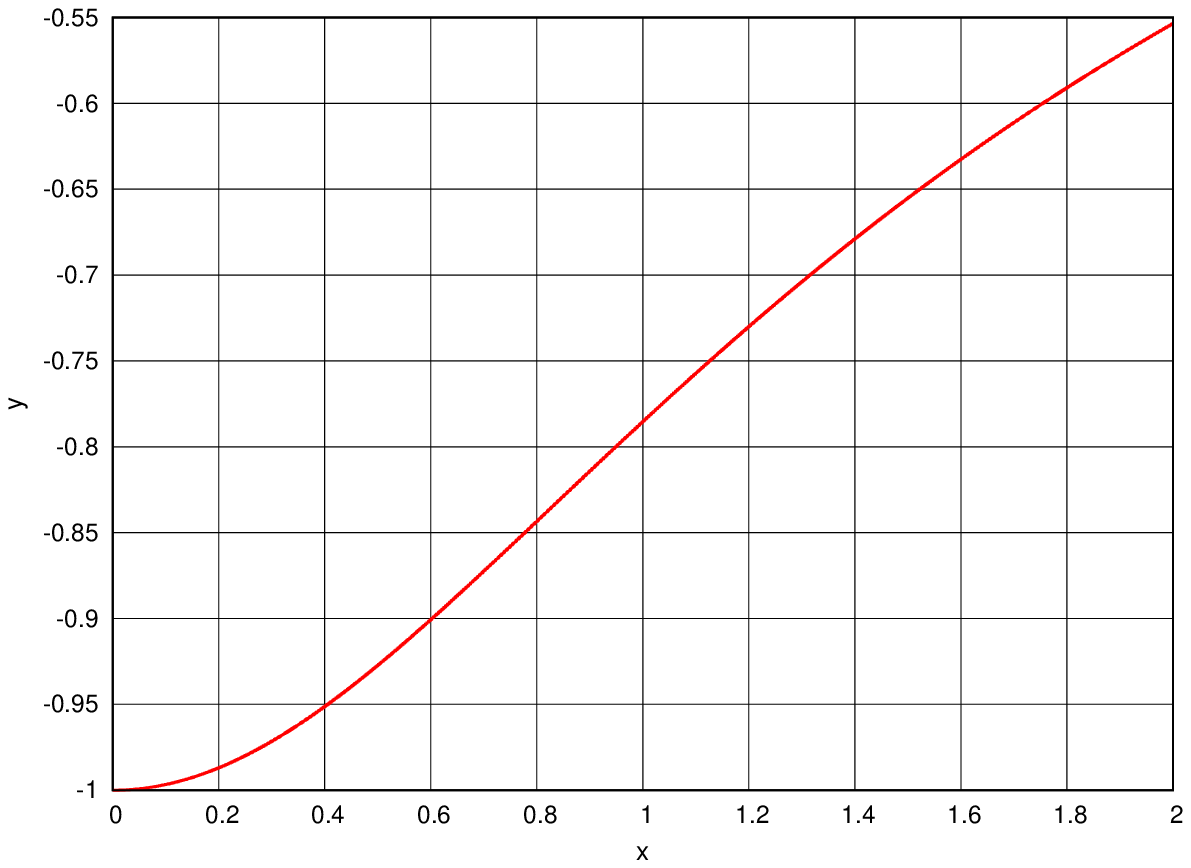}
\caption{$y=\arctan x/x\,\,\,(0\leq x \leq 2)$ }
\end{center}
\begin{center}
\includegraphics[width=90mm]{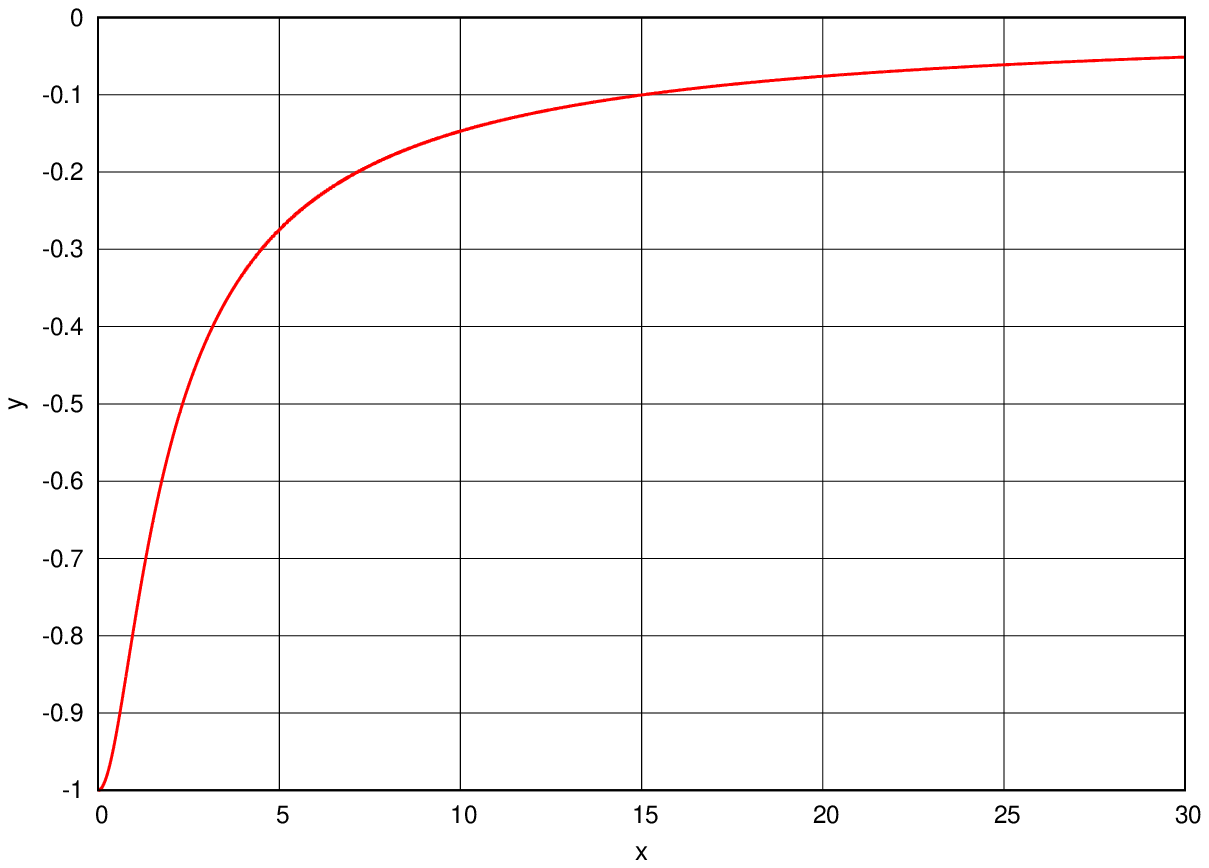}
\caption{$y=\arctan x/x\,\,\,(0\leq x \leq 50)$}
\end{center}
\end{figure}
Using $\arctan(x) = x-x^3/3+\cdots$ and $\lim_{x \rightarrow \infty} \arctan(x) = \pi/2$, 
 (\ref{eq:20}) can be approximated in each region as follows:
\begin{eqnarray}
\label{eq:21}
V(r)
&=&
\left \{
\begin{array}{ll}
\displaystyle - \frac{ze^2 m_c}{2\pi^2} + \frac{ze^2m_c^3}{6 \pi^2} r^2 & (m_c r \ll 1), \\
 & \\
\displaystyle - \frac{ze^2}{4\pi r}  & (m_c r \gg 1).
\end{array}
\right.
\end{eqnarray}
The modified Coulomb potential has interesting properties that it becomes the ordinary Coulomb potential at a long distance and becomes finite at $r = 0$.
We show the graph for each region in Figs. 1 and 2.

Let us evaluate the cutoff $m_c$.
The boundary $r_c$ of each region given by (\ref{eq:21}) satisfies $m_c r_c=1$, that is, $r_c=1/m_c$.
In the region which is larger than the nucleus radius $r_N= 10^{-15}$ m, Rutherford's scattering experiment suggests that the Coulomb potential is the ordinary one.
Therefore we find $r_c \leq r_N$. That is,
\begin{equation}
\label{eq:22}
m_c \geq 1/r_N = 200 \mbox{ MeV},
\end{equation}
where we used $\hbar c = 197\times10^{-15}$ MeV$\cdot$m.

\section{The Lamb shift}
Since the potential around $r=0$ has changed,
the $2s_{1/2}$ electron contributes to the Lamb shift.
The second term $e^2m_c^3/(6 \pi^2) r^2$
of (\ref{eq:21}) for $m_c r \ll 1$ generates an energy of a one-dimensional harmonic oscillator, $(n_h+1/2)\omega_h$, 
where $\omega_h \equiv \sqrt{e^2 m_c^3/(3\pi^2 m_e)}$, and $n_h$ is the quantum number of the one-dimensional harmonic oscillator.
Therefore, the energy change near $r = 0$ is
\begin{equation}
\label{eq:23}
E_{old} =-\frac{ e^2}{4\pi r}
\rightarrow
E_{new} = - \frac{e^2 m_c}{2\pi^2} 
+ \left. \left(n_h+\frac{1}{2} \right) \omega_h \right |_{n_h=0},
\end{equation}
where we chose the stable ground state $n_h = 0$.
When $r_c \ll a_0$,
the existence probability of the $2s_{1/2}$ electron in the sphere of radius $r_c$ is approximately $\int_0^{r_c} 4\pi r^2 dr |\psi_{20}(0) |^2$,
where $\psi_{nl}$ are the usual hydrogen atom wave functions.
Therefore the Lamb shift is 
\begin{eqnarray}
\label{eq:24}
\Delta E 
&\simeq&
\int_0^{r_c} 4\pi r^2 dr |\psi_{20}(0) |^2 (E_{new} - E_{old})
\nonumber \\
&=&
\int_0^{r_c} 4\pi r^2 dr \frac{1}{\pi} \left( \frac{m_e \alpha}{2} \right)^3 
\left \{ 
\left( - \frac{e^2 m_c}{2\pi^2} +\frac{1}{2}\omega_h\right)
-
\left( - \frac{e^2}{4\pi r} \right)
\right \}
\nonumber \\
&=&
\frac{\alpha^4 m_e^3}{2m_c^2} \left(\frac{1}{2}-\frac{2}{3\pi} \right)
+\frac{ \alpha^3 m_e^2}{6 m_c}\sqrt{\frac{\alpha m_e}{3\pi  m_c}}.
\end{eqnarray}
Substituting (\ref{eq:22}) into (\ref{eq:24}),
we obtain the contribution
\begin{equation}
\label{eq:25}
\Delta E \leq 0.1191\times10^{-6} \mbox{ eV}=28.8\mbox{ MHz}
\end{equation}
to the total Lamb shift of $1057$ MHz.
If $m_c=18.42$ MeV, we have $\Delta E = 1057$ MHz.
Interestingly, (\ref{eq:1}) with the cutoff $m_e$ replaced by 18.42 MeV agrees well with the experimental value:
\begin{equation}
\label{eq:26}
\Delta E = \frac{\alpha^5 m_e}{6\pi} \int^{18.42}_{1/a_0} \frac{d p}{p}
=1087 \mbox{ MHz}.
\end{equation}

\section{Zero-point energy density}
The canonical distribution assumption in this paper means that the energy created by virtual photons themselves is finite.
Let us calculate the zero-point energy density of the electromagnetic field.
The electromagnetic field operators with the probability density in the Coulomb gauge are 
\begin{equation}
\label{eq:27}
A_i(x) 
=
\int \frac{d^3p} { \sqrt{(2\pi)^3 2\omega(\boldsymbol{p})}} 
\sum_{\lambda=1}^2 \varepsilon_i^\lambda(\boldsymbol{p})
\{
 a^{\lambda}_{\boldsymbol{p}}e^{-ipx} 
+
a^{\lambda\dagger}_{\boldsymbol{p}} e^{ipx}
\}
 \rho(\boldsymbol{p}, \bar{u})^{1/2},
\end{equation}
where the coefficients of the expansion satisfy the canonical commutation relations:
\begin{equation}
\label{eq:28}
[a_{\boldsymbol{p}}^\lambda, a_{\boldsymbol{q}}^{\lambda'\dagger}] = \delta_{\lambda\lambda'}\delta^{3}(\boldsymbol{p}-\mbox{\boldmath $q$}), \,\,\,\,
[a_{\boldsymbol{p}}^\lambda, a_{\boldsymbol{q}}^{\lambda'}] = 
[a_{\boldsymbol{p}}^{\lambda\dagger}, a_{\boldsymbol{q}}^{\lambda'\dagger}]
 = 0.
\end{equation}
Using (\ref{eq:27}), the Hamiltonian is calculated as
\begin{eqnarray}
\label{eq:29}
H 
&=& 
\int d^3 x \{  (\dot{A_j}(x))^2 - (-\frac{1}{4}F_{\mu\nu}F^{\mu\nu}) \} \nonumber \\
&=& 
\sum_{\lambda=1}^{2}
\int d^3p   \omega(\boldsymbol{p})  
\{
 a^{\lambda\dagger}_{\boldsymbol{p}} a_{\boldsymbol{p}}^\lambda 
+\delta(\boldsymbol{p}=0) 
\}
\frac{ e^{-|\bar{u}_\mu p^\mu(\boldsymbol{p})|/m_c } }{Z}
 \nonumber \\
&=& 
\sum_{\lambda=1}^{2}
\int d^3p   \omega(\boldsymbol{p})  
\left \{ 
a^{\lambda\dagger}_{\boldsymbol{p}} a_{\boldsymbol{p}}^\lambda 
+\frac{1}{2} \frac{V_\infty }{ (2\pi)^3}
\right \}
\frac{ e^{-|\bar{u}_\mu p^\mu(\boldsymbol{p})|/m_c } }{Z},
 \nonumber \\
\end{eqnarray}
where $V_\infty \equiv \int_{-\infty}^\infty d^3 x$, and 
we used the identities $\delta(\mbox{\boldmath $p$}=0) = \int d^3 x e^{i\boldsymbol{p} \cdot \boldsymbol{x}  }/(2\pi)^3 |_{\boldsymbol{p} \rightarrow 0} 
=\int d^3 x/(2\pi)^3=V_\infty/(2\pi)^3 $. 
Using (\ref{eq:29}), we have
\begin{eqnarray}
\label{eq:30}
\frac{<0'|H|0'>}{V_\infty}
&=& 
\sum_{\lambda=1}^{2}
\int_{-\infty}^{\infty} \frac{d^3p }{ (2\pi)^3} 
\frac{1}{2}\omega(\boldsymbol{p})
e^{-|\bar{u}_\mu p^\mu(\boldsymbol{p})|/m_c } \nonumber \\
&=& 
\int_{-\infty}^{\infty} \frac{d^3p }{ (2\pi)^3}  \omega(\boldsymbol{p}) e^{-\omega(\boldsymbol{p})/m_c } \nonumber \\
&=& 
\frac{3m_c^4}{\pi^2},
\end{eqnarray}
where we used (\ref{eq:12}) and (\ref{eq:14}). 
Thus, we obtained  a finite zero-point energy density.
Notice that the result is independent of the average of the four-velocities of real particles.

Substituting (\ref{eq:25}) into (\ref{eq:30}),
the zero-point energy density is
\begin{equation}
\label{eq:31}
\frac{<0'|H|0'>}{V_\infty}
=
\frac{3m_c^4}{\pi^2}
\geq 61 \mbox{ MeV}/\mbox{fm}^3,
\end{equation}
which is extremely large compared to the experimental value.
We propose one method to reduce this value.
According to statistical mechanics, $m_c$ in (\ref{eq:9}) is $kT$, where $k$ is the Boltzmann constant.
If the vacuum temperature is lower than the electromagnetic field temperature and close to zero,
the zero-point energy density becomes
\begin{equation}
\label{eq:32}
\frac{<0'|H|0'>}{V_\infty}
=
\left.
\frac{3(kT)^4}{\pi^2}
\right |_{T\simeq 0}
\simeq
0.
\end{equation}
Thus, we can obtain a small zero-point energy density.

\section{Conclusion}
In this study, we introduced  a relativistic canonical distribution $ e^{-|\bar{u}_\mu p^\mu(\boldsymbol{p})|/m_c } /Z$ into the quantum field theory of the electromagnetic field. 
Here $\bar{u}$ is the average of the four-velocities of the particles created by the electromagnetic field.
$m_c$ has the meaning of a cutoff in ordinary quantum field theory.
As a result, we demonstrated the Coulomb potential is modified to $V(r)=- ze^2\arctan (m_c r)/(2\pi^2 r)$.
We have also shown that if the cutoff value $m_c$ is 18.42 MeV, the potential gives the Lamb shift 1057 MHz of the $2s_{1/2}$ electron.

The zero-point energy density of the electromagnetic field has become finite thanks to the Boltzmann factor $e^{-|\bar{u}_\mu p^\mu(\boldsymbol{p})|/m_c }$.
However, the cutoff 18.42 MeV gives the zero-point energy density that is larger than the experimental value.
To solve this, we introduced temperatures of quantum fluctuations, and we proposed an idea that 
the temperature of the vacuum is lower than that of the electromagnetic field and close to zero.

\end{document}